\documentstyle[preprint,aps,epsfig]{revtex}
\tightenlines
\begin{document}
\draft
\title{
\begin{flushright}
{OUNP-99-05\\
PDK-730}
\end{flushright}
{The Observation of a Shadow of the Moon in the Underground
Muon Flux in the Soudan 2 Detector.}\\
{\it The Soudan 2 Collaboration}
}
\author{
J.~H.~Cobb$^3$,
M.~L.~Marshak$^2$, 
W.~W.~M.~Allison$^3$,
G.~J.~Alner$^4$,
D.~S.~Ayres$^1$,
W.~L.~Barrett$^6$,
C.~Bode$^2$,
P.~M.~Border$^2$,
C.~B.~Brooks$^3$,
R.~J.~Cotton$^4$,
H.~Courant$^2$, 
D.~M.~Demuth$^2$, 
T.~H.~Fields$^1$,
H.~R.~Gallagher$^3$,
M.~C.~Goodman$^1$, 
R.~Gran$^2$, 
T.~Joffe-Minor$^1$,
T.~Kafka$^5$, 
S.~M.~S.~Kasahara$^2$,
W.~Leeson$^1$, 
P.~J.~Litchfield$^4$, 
N.~P.~Longley$^2$,
W.~A.~Mann$^5$, 
R.~H.~Milburn$^5$, 
W.~H.~Miller$^2$, 
C.~Moon$^2$,
L.~Mualem$^2$,
A.~Napier$^5$, 
W.~P.~Oliver$^5$, 
G.~F.~Pearce$^4$,  
E.~A.~Peterson$^2$,
D.~A.~Petyt$^4$,
L.~E.~Price$^1$, 
K.~Ruddick$^2$, 
M.~Sanchez$^5$,  
P.~Sankey$^2$,
J.~Schneps$^5$, 
M.~H.~Schub$^2$,  
R.~Seidlein$^1$, 
A.~Stassinakis$^3$, 
J.~L.~Thron$^1$,
V.~Vassiliev$^2$, 
G.~Villaume$^2$, 
S.~P.~Wakely$^2$,
N.~West$^3$,
D.~Wall$^5$ 
}

\address{$^1$Argonne National Laboratory, Argonne, IL 60439, USA }
\address{$^2$University of Minnesota, Minneapolis, MN 55455, USA }
\address{$^3$Department of Physics, University of Oxford, Oxford OX1 3RH, UK }
\address{$^4$Rutherford Appleton Laboratory, Chilton, Didcot,
Oxfordshire OX11 0QX, UK }
\address{$^5$Tufts University, Medford, MA 02155, USA }
\address{$^6$Western Washington University, Bellingham, WA 98225, USA }
\date{\today}
%
%
%
\maketitle
%


\begin{center}
Submitted to Physical Review D
\end{center}

\newcommand{\degree}{^{\circ}}      
\newcommand{\etal}{{\it et al.}}    
\newcommand{\thew}{\theta_{\rm EW}} 
\newcommand{\thns}{\theta_{\rm NS}} 
\newcommand{\var}{\mbox {\rm var}}  

\newpage
\begin{abstract}
A shadow of the moon, with a statistical significance of $5 \sigma$, has been
observed in the underground muon flux at a depth of 2090~mwe 
using the Soudan~2 detector.
The angular resolution of the detector is well described by a Gaussian with a
sigma $\le 0.3\degree$. 
The position of the shadow confirms that the alignment of 
the detector is known to better than $0.15\degree$ and has remained stable
during ten years of data taking.

\end{abstract}


\newpage

\section{Introduction}\label{sec:intro}

The Soudan 2 detector is a 963 tonne iron and drift-tube sampling 
calorimeter, 
designed to search for nucleon decay \cite{allison,allison2},
situated in the Soudan mine in northern Minnesota, USA.
The spatial resolution of the detector is $< 1$ cm in three dimensions.
Throughgoing muons are typically reconstructed with track lengths of
a few metres and the angular resolution for such tracks is expected to be 
a fraction of a degree. Because of the
good angular resolution and the uniform exposure of the detector 
over the period of ten years from January 1989, the detector is
excellently suited to a study of the possible existence of point 
sources of underground muons. 

A prerequisite for 
such a search is a thorough understanding of the alignment and 
angular resolution (point spread function or {\it psf\/}) of the 
detector. The angular diameters of the moon and sun are both 
$\sim 0.5^{\circ}$. Both objects will occlude the high energy cosmic ray 
primaries responsible for the underground muon flux. Since their 
angular sizes are small and comparable with the psf of the detector 
the observation of shadows in the underground muon flux will confirm both
the alignment and the angular resolution of the detector.

The potential observation of moon and sun shadows with
high energy cosmic rays was originally suggested by Clark 
\cite{clark}; the observation of the effect of the solar 
magnetic field on the cosmic ray shadow
of the sun has been suggested as a means of determining the 
composition of primary cosmic rays with energies of 
$\sim$100~TeV\cite{evans}. Shadows of the moon and sun have been 
observed in large EAS arrays \cite{cygnus,casa,hegra,tibet1}. One 
experiment, sensitive to primaries with a median energy of 
17~TeV, has observed a time-dependent displacement of the shadow of 
the sun which was attributed to the interplanetary magnetic 
field\cite{tibet2,tibet3}. The observation of a moon shadow in the 
underground muon flux has recently been reported \cite{macro1,macro2}. 
The 
data reported here were accumulated over a period of ten years 
during which the solar and interplanetary magnetic
fields varied considerably and results are presented only for the shadow
of the moon. The results of an analysis of the data for the shadow of 
the sun will be presented in a separate publication \cite{sunpaper}.

A shadow of the moon in the underground muon flux 
should be observable  if the transverse momentum impulse, $\Delta {\vec{p_t}}$,
imparted to a primary by the geomagnetic field is 
sufficiently small compared with the 
momentum of the primary.
The Soudan 2 detector is located at a depth of 2090~mwe at latitude 
$47.82\degree$~N, longitude $92.24\degree$~W in N\/E\/ Minnesota, USA. 
The local overburden is flat. 
The magnitude of the local geomagnetic field is 59~$\mu$Tesla and, 
because Soudan is close to the geomagnetic pole, the field at 
the detector is inclined at $14.9 \degree$ to the vertical, points down and to
the north, and lies almost in the plane of the local meridian.  
An estimate, based on an impulse approximation and assuming a dipole
field, of the transverse momentum  imparted to 
a primary cosmic ray yields 
$\Delta {\vec {p_t}} \sim c R_{\oplus}{\vec \beta} \wedge {\vec B}$
where $R_{\oplus} = 6378$~km is the radius of the earth. 
When averaged over azimuth for primaries from 
directions close to the moon, $\Delta {\vec {p_t}}$ is estimated to be 
$\sim 25$~GeV/c to
the west. The $p_t$ impulses in the {\sc n -- s} and vertical
directions average to zero. 10~TeV primaries would therefore suffer 
mean deflections of $\sim 0.14\degree$ to the west, leading to their mean
directions reconstructed in the detector being apparently displaced to
the east.
The rms value of $\Delta{\vec{p_t}}$ is $50$~GeV/c 
which would produce a broadening
of the shadow of the moon of $\sim 0.28\degree$ for 10~TeV primaries. 
A more precise estimate of geomagnetic effects is described below.

\par
The position of the 
moon lies within approximately $\pm 5\degree$ of the ecliptic which is 
itself inclined at an angle of $23.5\degree$ to the equatorial plane. 
The declination of the moon therefore varies between $\pm 28.5\degree$
throughout the year and its minimum zenith angle at Soudan is 
$\sim 20\degree$. A muon arriving at the detector from a direction close to the 
moon must therefore penetrate at least 2200~mwe of 
rock to reach the detector, corresponding to a minimum surface 
energy of $\sim0.8$~TeV. The mean slant depth is $\sim 3000$~mwe and the 
median surface energy of muons arriving 
underground from this direction is estimated to be $\sim 1.5$~TeV. 
Since the energy of a muon in an air shower is approximately one 
tenth of the energy of the primary, most muons observed in the detector 
originate from primaries with energies of approximately 6 to 
15~TeV or greater when the moon is closer to the horizon. Geomagnetic
effects are therefore expected to be small (${\cal{O}} (0.1\degree)$), 
if not entirely negligible, compared with the expected angular resolution 
of a fraction of a degree.  

Two other effects will influence the position and size of any shadow observed in
the detector. Multiple Coulomb scattering of the muons in the rock 
overburden will broaden the psf of the detector and, since
an object with fixed celestial coordinates moves in  
detector coordinates during a day, any misalignment of the detector will
also contribute to a broadening of the psf as well as displacing the position 
of a shadow \cite{thomson}.

The broadening of any shadow depends upon the momentum 
spectrum of the primaries. The Monte Carlo simulation described in reference 
\cite{smsk} has been used to study the contributions to the broadening from
geomagnetic deflections and multiple Coulomb scattering.
A sample of 14,000 muons from cosmic ray primaries with initial directions 
within  $3\degree$ of the position the moon (uniformly sampled throughout the 
period of data taking) was generated using the 
{\sc hemas}\cite{hemas} cascade and  {\sc sibyll}\cite{sibyll} hadronic
interaction codes. The muons were tracked with {\sc geant} to the Soudan 
depth.

The mean energy of the primaries from these directions which resulted in a 
muon at the detector is 19~TeV. 
The impulse approximation described above, together with the energy of
the primary, is used to estimate the geomagnetic deflection of the 
primary.
The broadening due to multiple scattering (and the shower production
mechanism) was determined by calculating
the angle between the directions of the primary and the muon underground.
The angular distributions ($dN/{d\theta^2}$) due to geomagnetic
dispersion and multiple scattering are very 
non-Gaussian, being sharply peaked towards zero and having tails 
extending beyond 4 degrees squared. The distributions are summarised in 
table~\ref{tab:broad} by the 50-percentile and mean values of $\theta^2$. The
broadening due the geomagnetic field is small. Multiple scattering gives
a mean square broadening of 0.43 degrees squared although the distribution
is peaked much more closely towards zero than a Gaussian distribution with
$\sigma = 0.46\degree$ which has the same $<\theta^2>$. The moon itself may be 
effectively taken as a point object. 

The angular size of the moon which will be
deduced from the data will be characteristic of the psf 
of the detector for a point source where the psf folds together
the angular dispersion due to detector resolution and alignment,
geomagnetic dispersion and multiple Coulomb scattering. The
detector angular resolution is expected to be a few tenths of a degree from
considerations of track length and spatial resolution.

\section{Data sample}\label{sec:datasample}

The Soudan~2 detector is described fully in references \cite{allison}
and \cite{allison2}. It is composed of two layers of 
$1 \times 1.1 \times 2.5$~m$^3$ modules containing corrugated steel
plates interleaved with 1~m long, 15~mm diameter drift-tubes filled
with argon/15\% carbon dioxide at atmospheric pressure. The three
coordinates of a track crossing a tube are obtained from an array of
crossed anode wires and cathode pads situated at the ends of the tubes, 
and drift times. Pulse height, which is proportional to the ionisation
deposited, is also recorded for each hit.
The electron drift velocity is $\sim 0.6$~cm/$\mu$sec;
the maximum electron drift distance is 50~cm. 
The detector has been surveyed to $0.1\degree$ 
about all three axes which are the vertical and very 
close (within $0.3\degree$) to {\sc e -- w} and {\sc n -- s}.
   
The detector has operated continuously since January 1989. Its size 
increased steadily to reach its present size of 
14~m~$\times$~8~m~$\times$~5~m in late 1993. Data taking has been 
very stable with  down time for maintenance and installation confined
to daytimes. Muons are currently accumulated at a rate
of approximately 7.5 million per year. A total sample of 58.5 million 
muons has been accumulated during the ten years until December 1998.

Two complementary software algorithms are used to reconstruct tracks 
in the detector. The first, `{\sc fmr}', is a fast algorithm which 
assembles small, clean, track segments and ignores noisy regions due to
catastrophic energy losses from bremsstrahlung or pair production.
The second, `{\sc search}', associates hits in roads to form complete
tracks. Both algorithms work with hits in the anode-time and 
cathode-time projections before a full three-dimensional track is 
defined. The electron drift velocities used by the two algorithms
are independent and are corrected for changes in atmospheric pressure
(which can amount to a few percent) and gas composition. In most cases
a penetrating track will be reconstructed by both {\sc fmr} and {\sc search}.
The {\sc search} algorithm usually reconstructs a greater length of track
but has somewhat worse systematics than {\sc fmr}. 
The systematic difference between the parameters of tracks reconstructed 
by both {\sc search} and {\sc fmr} is $ < 0.05\degree$.
Both algorithms tend to be unreliable for tracks lying close to the 
principal planes of the detector.

Stringent cuts are applied to the reconstructed data to 
ensure a high quality sample. 
The data are taken in runs which last approximately one hour. Any run 
which shows an anomalously high trigger rate or rate of reconstructed
tracks, or other symptom of detector malfunction, is rejected.
These cuts reject approximately 6\% of the data runs. The track selection 
requires either a {\sc search} track longer than 3~m which lies no
closer than $10\degree$ to the principal planes of the detector or, if 
such a track is not found, an {\sc fmr} track between 1~m and 5~m long.
In either case the selected track is also required to have a 
pulse height per unit length (i.e. $dE/dX$ \/) consistent 
with a minimum ionising particle and to have reconstructed local angles 
such that it would have triggered the detector.
These cuts reduce the total sample of 58.5 million muons to a sample of 
33.5 million for further analysis, with the major loss being the requirement on
track length.  

\begin{figure}[ht]
\centerline{\epsfig{figure=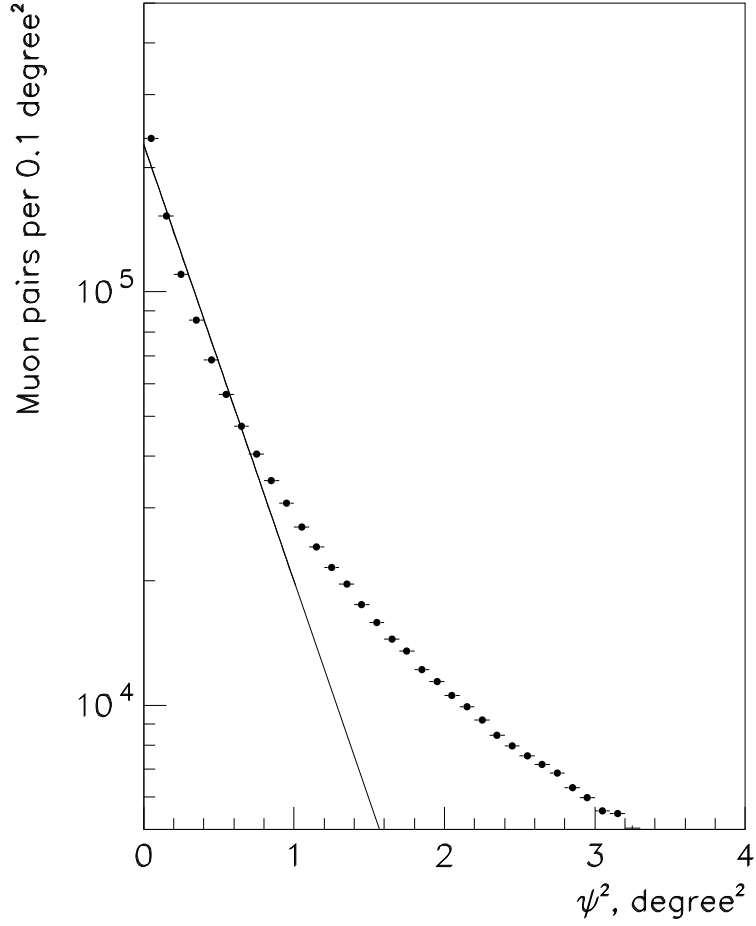,height=15cm,width=12cm}}
\caption 
{The distribution, $dN/d\psi^2$, of the square of the space angle 
between pairs of muons observed simultaneously in the detector. The 
line is the result of a fit 
$dN/d\psi^2 \propto \exp(-\psi^2/4 \sigma^2)$ 
in the region $\psi^2 < 1$ degree$^2$. The fitted value of $\sigma$ 
is $0.32 \degree$.}
\label{fig:dimuang}
\end{figure}

Because of its good spatial resolution, and the many 
(typically $\sim 100$) hits used to form tracks, 
the inherent angular resolution of the detector is expected to 
be a small fraction of a degree for tracks a few metres long. 
The angular resolution can be investigated by examining the 
distribution of space angle, $\psi$, between pairs of muons observed 
simultaneously (i.e. belonging to the same parent shower) in the 
detector. Figure~\ref{fig:dimuang} shows the distribution of $\psi^2$ 
for a sample of 1.26 million muon pairs. For a Gaussian psf 
$dN/d\psi^2 \propto \exp(-\psi^2/4\sigma^2)$ where $\sigma$ is the 
angular resolution of the detector\footnote{
$4\sigma^2$ appears in the denominator
because the distribution for pairs of muons results from the convolution of two
similar distributions.}. The distribution shown in 
Figure~\ref{fig:dimuang} can be described by a Gaussian with 
$\sigma = 0.32\degree$ for $\psi < 1 \degree$; the excess beyond
$1 \degree$ can be attributed to multiple scattering of the muons 
in the overburden. There is no evidence for any anisotropy in two dimensions.

\section{Data Analysis}\label{sec:data}

\subsection{One dimensional analysis}\label{sec:da1d}

\bigskip\noindent
The time ({\sc utc}) of each selected event is recorded with a 
clock synchronised 
to the {\sc wwvb} time standard and used to convert the direction of the 
reconstructed track
in detector coordinates to geocentric celestial coordinates 
(right ascension, $\alpha_\mu$, and declination, $\delta_\mu$ ). 
The celestial coordinates of the moon, $\alpha_m$ and $\delta_m$,
at the event time are also calculated. The moon's position
is corrected to topocentric coordinates using the latitude and longitude of 
the detector to account for the parallax due to 
the moon's relative proximity. The routines\cite{duffet} used to calculate the 
lunar coordinates are accurate to better than $0.01\degree$ over 
the entire period of the exposure; the parallax corrections amount to
as much as $\pm 0.7\degree$ and $-0.8\degree$ in right ascension 
and declination respectively.

\begin{figure}[htb]
\centerline{\epsfig{figure=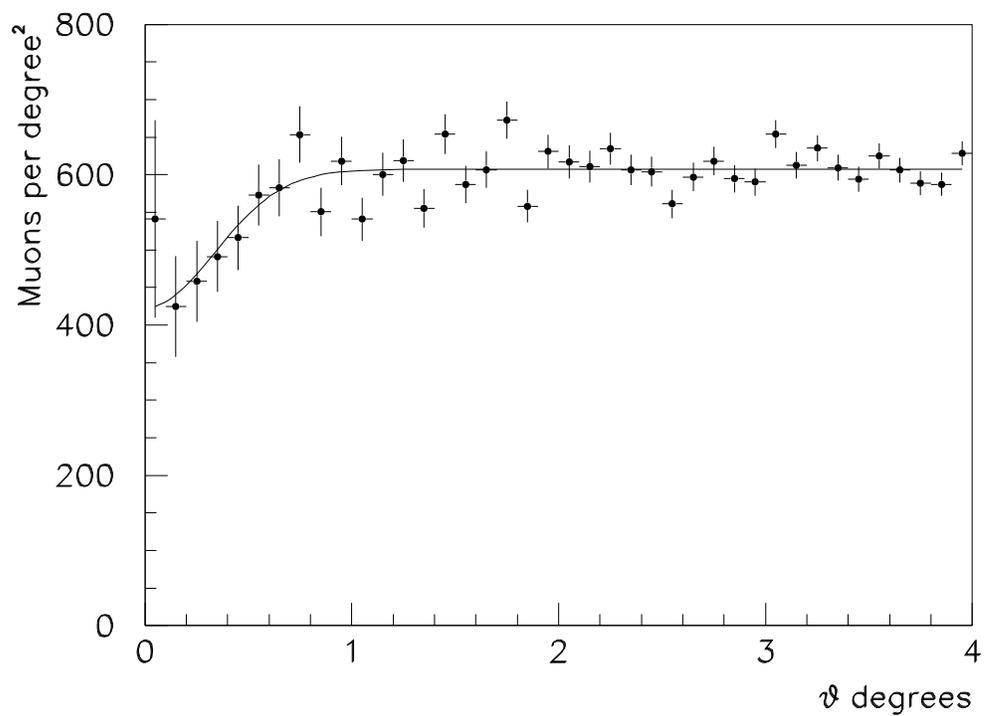,height=12cm,width=15cm}}
\caption 
{The angular density of muons, $(1/\pi)dN_\mu/d\theta^2$, versus $\theta$,
the angular distance between the muon direction and the nominal 
position of the centre of the moon. 
}
\label{fig:dndtsq}
\end{figure}

The space angle, $\theta$, between the direction of the muon track and 
the direction of the moon is histogrammed. The angular density of muons,
$dN_\mu/d\theta^2$, as a function of $\theta$ is derived from this histogram by
dividing the bin contents by the solid angle subtended by each bin.
Figure~\ref{fig:dndtsq} shows $(1/\pi) dN_\mu/d\theta^2$ versus $\theta$. In 
the
absence of a shadow this distribution should be flat. There is, however, 
a clear deficit of events in the first
few bins which is attributed to the shadowing effect of 
the moon. The
significance of the shadow is tested by comparing the difference in
$\chi^2$ between the best fit to a flat distribution ($\chi^2 = 82.9$)
and the best fit ($\chi^2 = 58.3$) to the form 
\begin{equation}
{dN_\mu\over{d\theta^2}} 
= \pi \lambda 
\left( 
1 - \pi(R_m^2/2 \sigma^2) \exp(-\theta^2/2\sigma^2)
\right)
\nonumber
\end{equation}
where $R_m = 0.26 \degree$ is the mean angular radius of the moon and
$\lambda$ and $\sigma$ are free parameters
representing the angular density of muons and the detector angular resolution
respectively. 
The fitted function treats the moon as a point object at $\theta = 0$ 
which removes $\lambda \pi R_m^2$ muons from the sample.
Numerical studies indicate the the effect of the finite angular 
size of the moon is relatively unimportant if the detector angular resolution
is of the order of $0.3\degree$. 
The best fit values of the parameters are 
$\lambda = 607.1 \pm 3.48$ muons per square degree and 
$\sigma = 0.333\degree \pm 0.048\degree$. The value of $\lambda$ implies
that a total of 128.9 muons (about one per month) are missing from the data 
sample due to the 
presence of the moon. The improvement in $\chi^2$ between 
the two fits of 24.6 for two degrees of freedom implies that chance that the 
observed deficit is a random fluctuation is $< 10^{-5}$: a shadow of the
moon is observed with a statistical significance of $5 \sigma$. 

Figure \ref{fig:dndtsq} clearly demonstrates the existence of a shadow
of the moon in the underground muon flux and indicates that the psf
of the detector is the expected small fraction of a degree.
There is no evidence for any large misalignment of the detector.

\subsection{Two dimensional analysis}\label{sec:da2d}
  
The procedure to extend the analysis to two dimensions is straightforward
and involves three stages. In the first, maps (two dimensional histograms) 
of the angular muon density in a region centred on the
moon are constructed from the arrival directions of the muons. In the
second, the maps are smoothed (rebinned) with a two-dimensional 
kernel to form a map of the local mean angular density of muons. The expected  
angular density in the absence of a shadow is estimated at the same time. 
Finally the expectation is subtracted from the smoothed data and $Z$,
a test statistic which represents the normalised excess or deficit at 
a point, is evaluated.

The celestial coordinates of a muon track are transformed into a 
coordinate system centred 
on the moon by first a rotation of $-\alpha_m$  about the polar axis
followed by a rotation of $-\delta_m$ around an axis 
in the equatorial plane. In this new system 
the $x$-axis points towards the moon  and the $y$-axis lies in 
the equatorial plane. The $x$ direction cosine, $c_x$, is therefore 
equal to $\cos \psi$ where $\psi$ is the angle between the track and 
the direction of the moon.  To a first approximation 
{\mbox {$c_y \sim (\alpha_\mu - \alpha_{m}) \cos \delta_{m}$}} and 
{\mbox {$c_z \sim \delta_\mu - \delta_{m} $}}. 
For tracks at small angles ($< 10\degree$) to the direction 
of the moon,  {\mbox {$1- c_x^2 = c_y^2 + c_z^2 \sim \psi^2 $}}, and
the space angle, in degrees, is well approximated by 
{\mbox {$\psi = ({180\degree/\pi})(c_y^2 + c_z^2)^{1\over 2}$}}.
Subsequent analysis is made using the quasi-angular coordinates 
{\mbox {$\thew = ({180\degree/\pi}) c_y$}}  and 
{\mbox {$\thns = ({180\degree/\pi}) c_z$}} which represent the angular 
distances between the direction of the muon and the moon in 
the west--east and south--north directions respectively. 

\begin{figure}[ht]
\centerline{\epsfig{figure=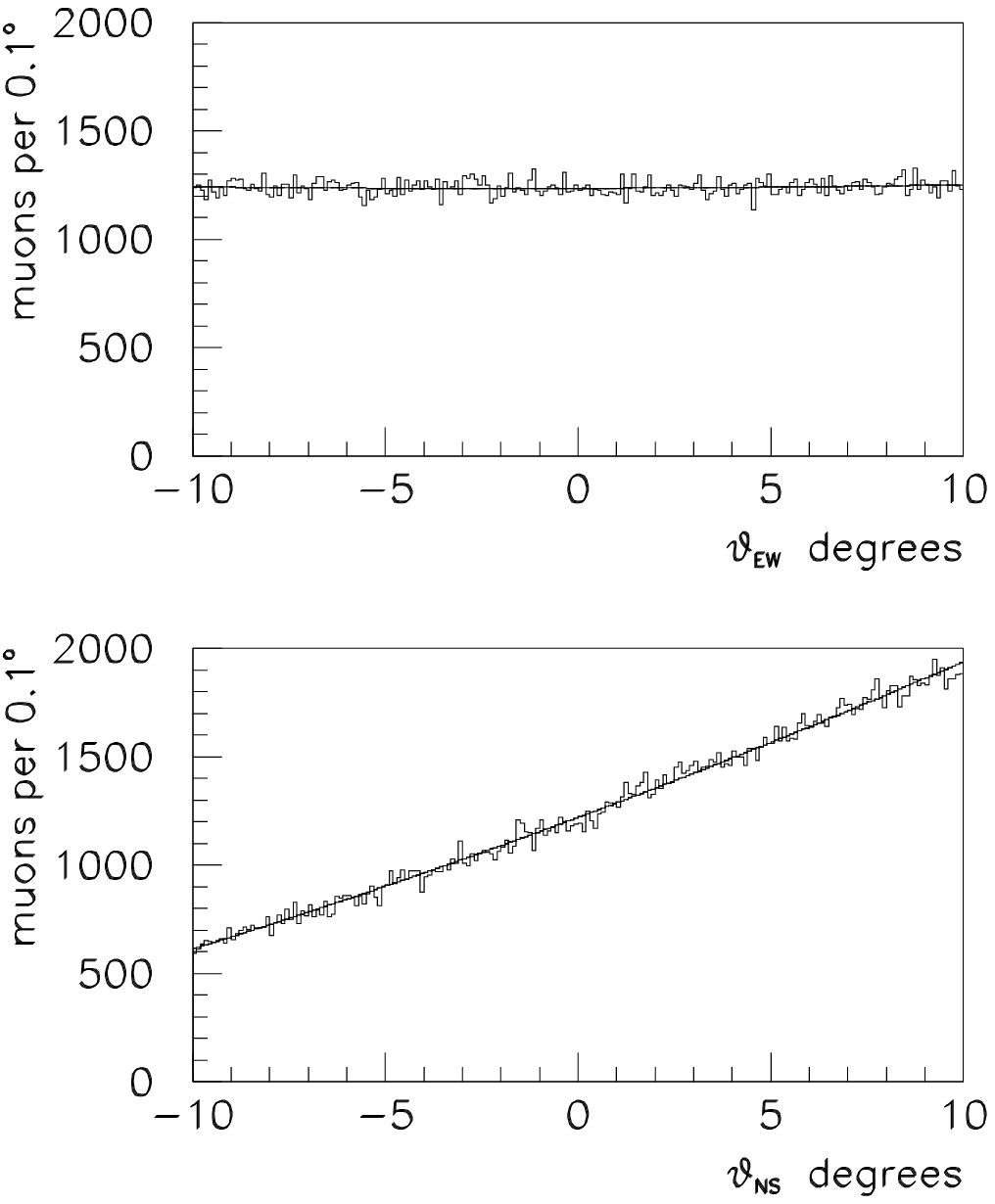,height=15cm,width=12cm}}
\caption 
{The $\thew$ and $\thns$ projections of the raw data in the  
$20 \degree \times 20 \degree$ low resolution map. }
\label{fig:moondata}
\end{figure}

Tracks pointing towards the moon, i.e. with $c_x >0 $, are 
selected. $2.44 \times 10^5$ tracks lie within 
an area of $\pm 10\degree \times \pm 10 \degree$ around the 
moon. The coordinates $\thew$ and $\thns$ of the muons are 
used to make low- and high-resolution maps (two-dimensional  distributions), 
covering $20\degree \times 20\degree$ and $8\degree \times 8\degree$ regions
with the moon at their centres. 
The bin sizes used are
$\Delta \thew \times \Delta \thns = 0.1\degree \times 0.1\degree$ and
$\Delta \thew \times \Delta \thns = 0.04\degree \times 0.04\degree$. 
The  $\thew$ and $\thns$ projections of the low resolution map are shown 
in Figure \ref{fig:moondata}. The flatness of the $\thew$ 
distributions is a result of the many cycles of averaging in the 
east-west (right ascension) direction during the  long exposure of 
the detector. The rise in the number of muons with 
$\thns$ (declination) results from the decreasing  overburden
for muons with small zenith angles. 
(The moon is never overhead at the detector latitude of 
$47.8\degree$.)

The density, $\lambda(\thew,\thns)$, of muons in the maps 
expected in the absence of a shadow is estimated by 
assuming that $\lambda$ factorises, in other words 
that the number of events expected in a bin centred at 
$\thew,\thns$ can be represented by
$dN_\mu = \lambda d\thew d\thns = Y(\thew) \times Z(\thns) d\thew d\thns$. The functions 
$Y(\thew)$ and $Z(\thns)$ are obtained by fitting quadratics  
to the $\thew$ and $\thns$ projections. 
The results of the fits for the low resolution map are shown in 
Figure \ref{fig:moondata}. The fits are good with 
$\chi^2 /{\mbox{\rm ndf}} = 189.1/198$
for the $\thns$ projection and  
$\chi^2 /{\mbox {\rm ndf}} = 170/199$ for the 
$\thew$ distribution.

The fitted functions $Y(\thew)$ and $Z(\thns)$  are used to predict the  
muon density, 
{\mbox {$\lambda(\thew , \thns) = d^2N_\mu/d\thew d\thns$}},
in the absence of a shadow 
at any point in the  map. As can be seen from Figure \ref{fig:moondata},
$\lambda$ varies by a factor of three over the range of $-10\degree$
to $+10\degree$ in $\thns$ direction but can be taken to be flat on a scale
of $1\degree$, allowing quantitative predictions to be made for the 
two-dimensional analysis which is described below. 
For the both the low- and
high-resolution maps, the $\chi^2$ between the number of events
in any bin, $n_{ij}$, and the prediction, 
$m_{ij} = Y(\thew) \times Z(\thns) \Delta\thew \Delta\thns$, is 
satisfactory with 
$\chi^2/{\mbox {\rm ndf}} = 40076.6/39994$ for the coarse map and 
$\chi^2/{\mbox {\rm ndf}} = 40559.4/39994$ for the fine map.

The predicted angular density of muons at the centre of the low resolution 
map is $ \lambda(0,0) = 608.3 $ muons per degree squared,
consistent with the angular density obtained from the fit to the
one-dimensional $dN_\mu/d\theta^2$ distribution.
The method used to estimate $\lambda$, however, includes the central region 
where there is a known deficit of events (caused by the shadow of the moon) and 
therefore results in a slight underestimate of the expectation.
The underestimate will be greater for the finer binned map; the predicted 
angular density of muons at the centre of the high resolution map is 
$ \lambda(0,0) = 592.1 $ muons per degree squared.
To avoid any bias caused by this underestimate, the value of $\lambda$ is taken 
uniquely from the low resolution map in the subsequent stages of analysis.

A two-dimensional Gaussian kernel of angular width $\sigma_k$ 
is used to rebin the data in the maps. The contents, $n_{ij}$, of 
bin $i,j$ are replaced with 
$d_{ij} = \sum_k\sum_l n_{(i+k)(j+l)} w_{kl}$. The weight 
$w_{kl} = (1/2\sigma_k^2)\exp (- \psi^2_{kl}/2\sigma_k^2) 
\Delta \thew \Delta \thns$ \/,
where $\psi_{kl}$ is the
angular distance between bin $i,j$ and bin $k,l$. The sums over $k$ 
and $l$ include all bins within $3\sigma_k$ of bin $i,j$ and 
$\sum_k\sum_l w_{kl} = 1$.

The expected number of events, $m_{ij}$, in bin $i,j$ in the absence of a 
shadow,  is predicted from the fits to the projections of the map 
described above. The resultant map of $m_{ij}$ is then treated identically to 
the raw data map to obtain the rebinned expected angular density $b_{ij}$. Maps 
are then made  of the normalised statistic 
$Z_{ij} = (d_{ij} - b_{ij}) / \sqrt{\var(b_{ij})}$ where 
$\var(b_{ij}) = \sum_k\sum_l m_{(i+k)(j+l)} w_{kl}^2$.
The value of $\lambda$ is sufficiently large that this
procedure, which treats the statistical fluctuations of the 
background as Gaussian, introduces negligible error despite the 
inherent Poisson fluctuations of the small number of events 
(typically eight) in any bin\cite{SiegRab}. 

Since the bin size of the maps is small compared with $\sigma$ 
(the angular resolution of the detector) and $\sigma_k$ (the 
rebinning kernel), and the background is flat over a region 
$ \gg \sigma$, the expected value of $Z$  for a point 
source (or sink) of strength $N_m$ may be estimated by treating the summations 
as integrals to yield 
\begin{eqnarray}
 Z  & = 
        & \left ( {\sigma_k^2 \over {\sigma_k^2 + \sigma^2}}\right ) 
          {N_m \over {\sigma_k \sqrt{\pi \lambda}}} 
\label{eq:zeq}
\end{eqnarray}
and
\begin{eqnarray}
 Z_m      & = & -\left ({\sigma_k^2 \over {\sigma_k^2 + \sigma^2}}\right ) 
          {R_m^2 {\sqrt{\pi \lambda}} \over {\sigma_k}} \,.
\label{eq:zeq2}
\end{eqnarray}
$Z$ is maximised when $\sigma_k = \sigma$, i.e. when the rebinning
kernel is matched to the angular resolution of the detector. 
Equation (\ref{eq:zeq2}) predicts $Z_m = -4.43$ for the shadow of the
moon using $\lambda (0,0) = 608.3$ muons per degree squared and
matching $\sigma_k$ to the value $\sigma = 0.333\degree$ found from the
fit to the $dN_\mu/d\theta^2$ distribution described in section \ref{sec:da1d}.

\begin{figure}[ht]
\centerline{\epsfig{figure=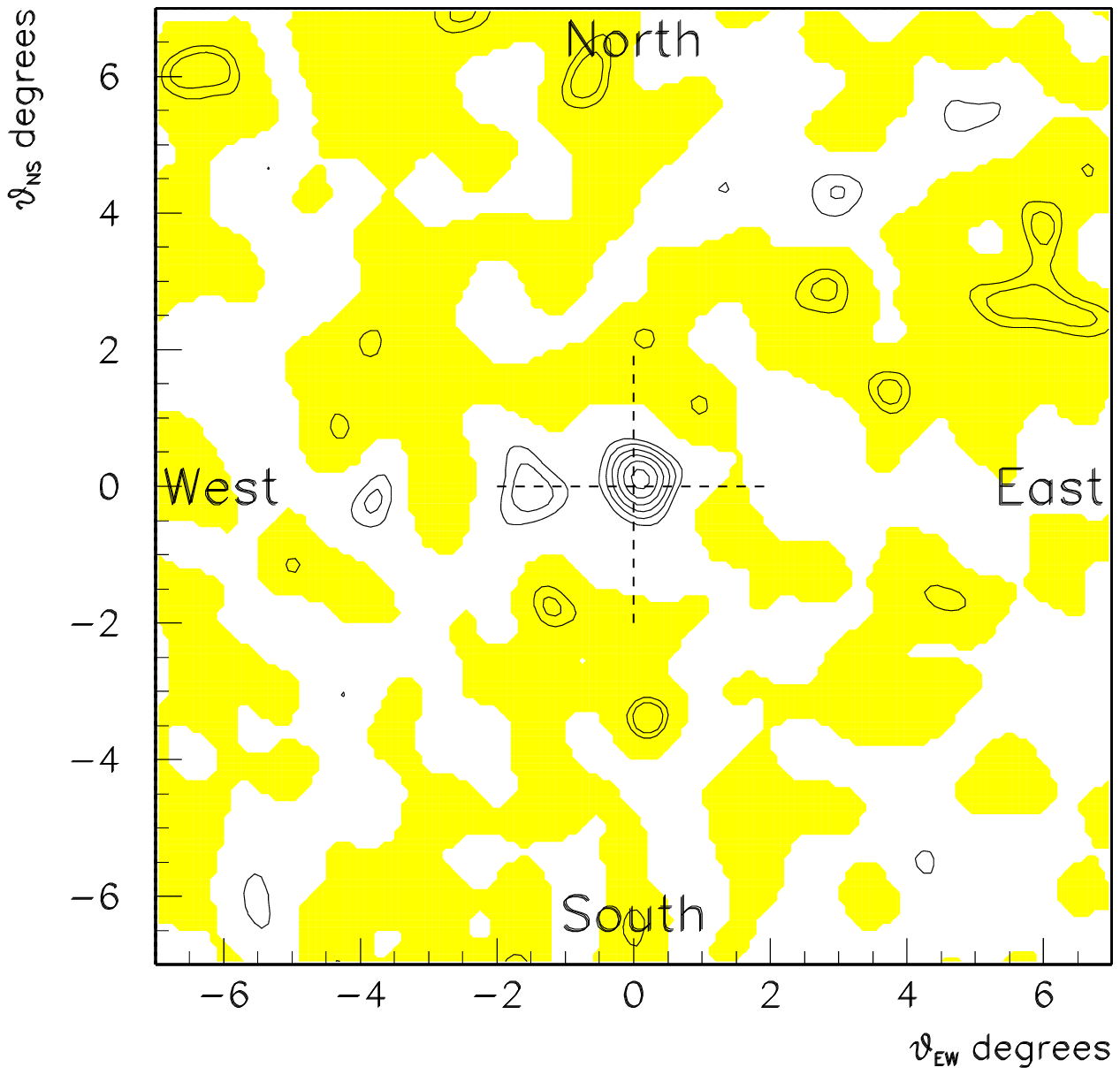,height=15cm,width=15cm}}
\caption 
{Contour map of normalised deviations, $Z$, for a 
$\pm 7 \degree \times \pm 7 \degree$ region 
centred on the moon made with a rebinning kernel with 
$\sigma_k = 0.333 \degree$. The contour lines are spaced by 
$\Delta Z = 0.5$ and shown only where $|Z| \ge 2.0$.
Regions with $Z > 0$ are shaded.}
\label{fig:moonzc}
\end{figure}

A contour map of normalised deviations, $Z$, made with a kernel size of
$\sigma_k = 0.333\degree$ for a 
$\pm 7\degree \times \pm 7\degree$ region centred on the moon
is shown in Figure \ref{fig:moonzc}. To avoid confusing detail,
contours are shown only for the regions where $|Z|\ge 2.0$. The shadow of
the moon is unambiguously visible as the deep minimum  
($Z = -4.69$) at the centre. The 25 extrema with $|Z| \ge 2.0$ in the 
entire area of 196 square degrees are consistent with the 31 expected 
from statistical fluctuations\cite{SiegWor}; there are no other minima
with $Z < - 3.0$. The observed value of $Z = -4.69$ for the shadow
suggests that the angular resolution is somewhat better 
than $0.333\degree$ determined from the one-dimensional analysis.

\begin{figure}[ht]
\centerline{\epsfig{figure=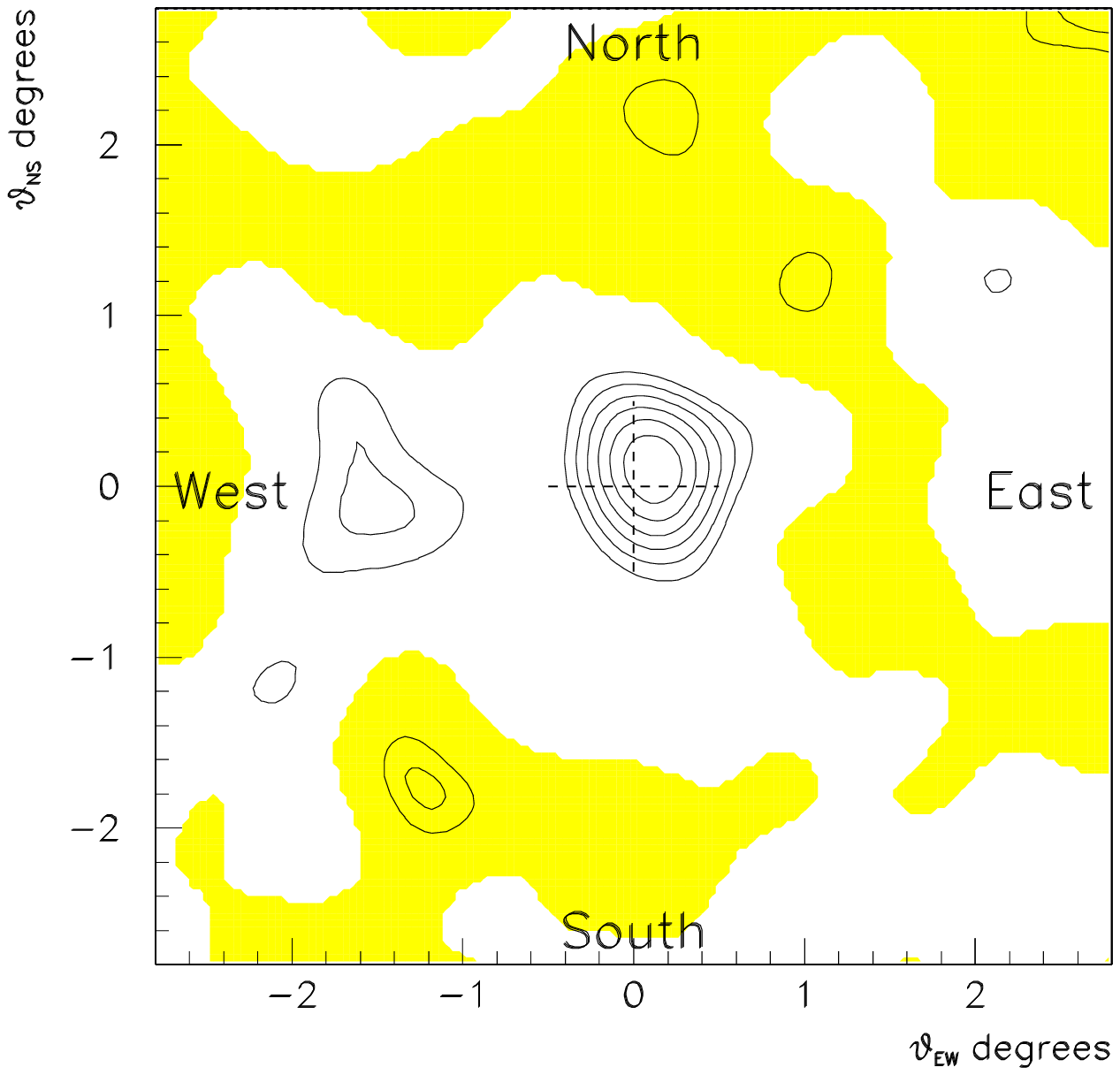,height=15cm,width=15cm}}
\caption 
{Contour map of normalised deviations, $Z$, for a 
$\pm 2.8 \degree \times \pm 2.8 \degree$ region 
centred on the moon made with a rebinning kernel with 
$\sigma_k = 0.29 \degree$. The contour lines are spaced by 
$\Delta Z = 0.5$ and shown only where $|Z| \ge 2.0$.
Regions with $Z > 0$ are shaded.}
\label{fig:moonzf}
\end{figure}

The bin size of $0.1\degree \times 0.1\degree$ is rather too 
coarse compared with
the angular resolution of approximately $0.3\degree$ to allow a precise
determination of the position of the shadow to be made. A further 
analysis is made using the finer binned map with a bin size of 
$0.04\degree \times 0.04\degree$.
Figure \ref{fig:moonzf} shows a contour map of $Z$ for a 
$\pm 2.8\degree \times \pm 2.8\degree$ region 
using a kernel size of $\sigma_k = 0.29\degree$ which was 
chosen to minimise $Z$ for the shadow. In this map $\lambda$ has
been evaluated using the parameters determined from the fits for
$Y(\thew)$ and $Z(\thns)$ to the coarser map to avoid the 
underestimate discussed above. 

With the finer binning, the minimum, which has a depth $Z=-4.98$,
at ($\thew = 0.1\degree,\thns=0.1\degree$) is within $0.15\degree$ 
of the nominal position of the moon. The $Z=-4.5$ contour, which
corresponds to $\sim75$~\% {\sc cl}, encompasses the origin and there is 
therefore no evidence that the detector is incorrectly aligned. 
It is not possible to separate what fraction of the small observed
displacement of the shadow could be due to misalignment, the statistics of the
sample or geomagnetic effects, although the 
displacement of $0.1\degree$ east is consistent with the estimated
geomagnetic displacement of $0.076\degree$ discussed in 
section~\ref{sec:intro}.

\section{Long term stability of the Soudan~2 detector}

As described in section \ref{sec:intro} the Soudan~2 detector was assembled
over a period of five years ending in 1993. Some modules were removed from the
detector and refurbished in 1994 and 1995. The detector continued to operate
during these periods and was re-surveyed after each movement.  
The clean shadow of the moon demonstrated in sections \ref{sec:da1d} and 
\ref{sec:da2d}, and the statistics of the 
sample allow the data to be subdivided to check the stability of 
the alignment of the detector.

\begin{figure}[ht]
\centerline{\epsfig{figure=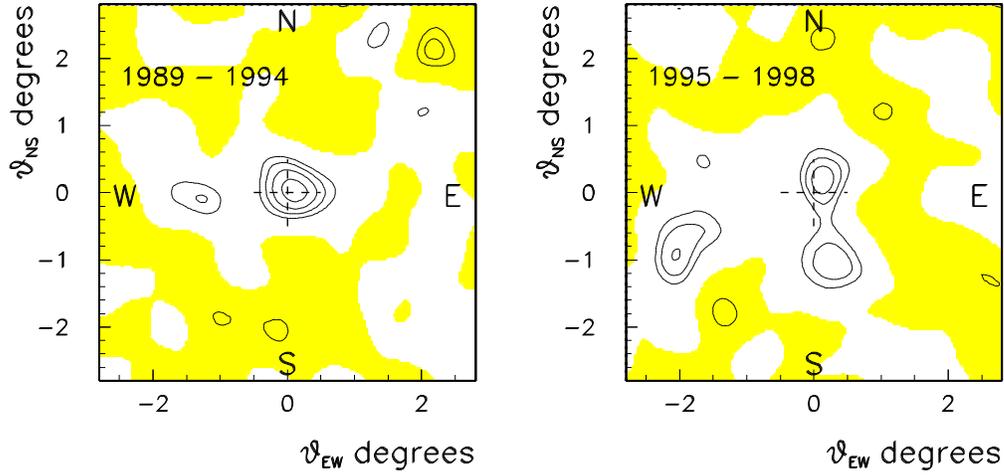,height=9cm,width=16cm}}
\caption 
{Contour maps of normalised deviations, $Z$, for a 
$\pm 2.8 \degree \times \pm 2.8 \degree$ region 
centred on the moon made with  
$\sigma_k = 0.29 \degree$ for the periods 1989 -- 1994 (left)
and 1995 -- 1998 (right). The contour lines are spaced by 
$\Delta Z = 0.5$ and shown only where $|Z| \ge 2.0$.
Regions with $Z>0$ are shaded.}
\label{fig:moonzf2}
\end{figure}

The data have been divided into two approximately equal samples for the periods
of 1989 to 1994 inclusive, and 1995 to 1998. The analysis described in
section \ref{sec:da2d} has been repeated for each period. The high 
resolution maps, made with $\sigma_k = 0.29\degree$,
for both periods are shown in Figure \ref{fig:moonzf2}. A shadow of the moon is
clearly seen in each map. The positions and depths of the shadows are
given in table \ref{tab:perz}. The depths of the minima are both close to the
expected value of $Z = - 4.98/\sqrt{2} = -3.52$ expected for half the total
sample. The positions of these shadows confirm that the alignment of the 
detector has remained stable during the ten year period.

\section{Conclusions}

An shadow of the moon in the underground muon flux has been 
observed using the Soudan~2 detector with a statistical significance of 
$5 \sigma$.
The size and position of the shadow confirm that the alignment of the detector 
is correct to better than $0.15 \degree$ and that the psf
of the detector can be adequately described by a Gaussian 
with $\sigma = 0.29 \degree$. 
The geomagnetic deflection of the primaries, which is expected to 
be of the order of $0.1\degree$ 
prohibits a more precise statement about the accuracy of
the alignment of the detector. The subdivision of the data into two periods
confirms that the alignment of the detector has remained stable over the 
ten years from 1989 to 1998.
These data confirm the excellent capabilities of Soudan~2 as a 
detector of possible astrophysical point sources of underground muons.

\section*{Acknowledgements}

This work was undertaken with the support of 
the U.S. Department of Energy, the State and University of
Minnesota and the U.K. Particle Physics and
Astronomy Research Council.

We wish to thank the following for their help with the experiment: 
the staffs of the collaborating laboratories; the
Minnesota Department of Natural Resources for allowing use of the 
facilities of the Soudan Underground Mine State Park; the staff of 
the Park, particularly Park Managers D. Logan and P. Wannarka, for 
their day to day support, and B. Anderson, J. Beaty, 
G. Benson, D. Carlson, J. Eininger and J. Meier of the
Soudan Mine Crew for their efficient running of the experiment.

We also wish to thank Prof. D. O. Siegmund of the Department of 
Statistics, Stanford University, for his help, advice and interest 
in our method of analysing the data using Gaussian smoothing.

\newpage

\begin{table}[hb]
\begin{center}
\begin{tabular}{lccc}                                                        
Source of broadening & $\theta^2_{50}$    & $<\theta^2>$    & $<\theta>$  \\
                     & (degrees$^2$)      &  (degrees$^2$)  &  (degrees)  \\
Moon (flat disc)     & 0.034              & 0.034           &  --         \\
Geomagnetic          & 0.04               & 0.32            &  0.076 west \\
Multiple scattering  & 0.095              & 0.43            &  0.0        \\
Gaussian ($\sigma = 0.46\degree$)            
                     & 0.3                & 0.43            &  --         \\
\end{tabular}
\caption{ Calculated contributions to the broadening of a shadow of the
moon. The first column gives the 50-percentile of the distribution; the 
last row gives the values for a Gaussian distribution for comparison. The
last column gives the mean displacement of a particle trajectory; 
the reconstructed trajectory, and hence any shadow, is displaced in the
opposite sense.}

\label{tab:broad}
\end{center}
\end{table}

\begin{table}[hb]
\begin{center}
\begin{tabular}{lllll}
Period       & $\lambda(0,0)$ & $Z_{\min}$  &$\thew$       & $\thns       $   \\
1989 -- 1998 & 608.3          & -4.98       & $0.1\degree$ & $0.10\degree $   \\
1989 -- 1994 & 295.9          & -3.78       & $0.1\degree$ & $0.02\degree $   \\
1995 -- 1998 & 311.8          & -3.39       & $0.1\degree$ & $0.18\degree $   \\
\end{tabular}
\caption{ Observed positions of the shadow of the moon for the entire data
sample, and the periods 1989 -- 1994 and 1995 -- 1998.
}
\label{tab:perz}
\end{center}
\end{table}
\newpage
\newpage
\newpage
\newpage

\newpage

\end{document}